White paper:

# Accounting for carbon emissions caused by cryptocurrency and token systems

*Crypto Carbon Ratings Institute (CCRI)[1]*

*https://carbon-ratings.com*

*Ulrich Gallersdörfer, Lena Klaaßen, Christian Stoll*

*Original Version: November 2021*

*Version 2: March 2022*

*Version 3: March 2023*

---

[1] *Crypto Carbon Ratings Institute (CCRI)* is a research-driven company providing data on sustainability aspects of cryptocurrencies, blockchain and other technologies. CCRI was founded by Ulrich Gallersdörfer, Lena Klaaßen and Christian Stoll in 2021. More information: https://carbon-ratings.com.

**Introduction**

The electricity consumption and related carbon emissions of cryptocurrencies such as Bitcoin are subject to extensive discussion in public, academia, and industry. Various estimations have compared Bitcoin's carbon footprint to the emissions produced by countries over the past years (Stoll et al., 2019).[2] As cryptocurrencies continue their journey into mainstream finance, incentives to participate in the networks and consume energy to do so remain significant.

As climate risks and carbon exposure of companies increasingly move into focus (Krueger et al., 2020), stakeholders along the crypto value chain are facing challenges on how to allocate the share of carbon emissions they need to account for. So far, the focus of research has mainly been on calculating the electricity consumption and carbon footprint of the entire network of a cryptocurrency. First guidance on how to allocate the carbon footprint of the Bitcoin network to single investors has been proposed (de Vries et al., 2021)[3], however, a holistic framework capturing a wider range of cryptocurrencies and tokens remained absent until CCRI and Southpole jointly published a report putting forth an approach to allocate crypto-related emissions across the value chain to holdings as well as transactions (CCRI and SouthPole, 2022).

This white paper builds on CCRI and South Pole (2022) by providing detailed guidance on how to allocate emissions caused by cryptocurrencies and tokens in different kinds of networks. In general, any activity within a network, either holdings or transactions, can be the cause of emissions. Based on the consensus mechanism of the blockchain network, different drivers of emissions can be identified; depending on the respective share of these drivers, emissions can be allocated respectively. Following CCRI and South Pole (2022), we coin this approach as "hybrid allocation".

For *Proof of Work (PoW) networks*, we propose the combination of two key drivers of electricity consumption (and carbon emissions, respectively), namely block rewards and transaction fees. This hybrid approach combines two approaches (holding-based and transaction-based allocation) and allocates emissions based on holding and transaction volumes. This allows capturing the specificities of many different PoW cryptocurrencies and their tokens with one consistent approach, by accounting for their underlying mining revenue structures. It can be applied to applications running on these networks as well as second-layer approaches such as the Lightning network.

For *Proof of Stake (PoS) networks*, we propose the differentiation between sources of electricity consumption (and carbon emissions, respectively). Given the extensive research and empirical data collection through measurements that have been conducted for Proof of Stake networks (CCRI, 2022b, 2022c, 2022d, 2022e)[4], we may identify marginal electricity consumption of transactions as well as the base electricity consumption of hardware without any transactions which can be allocated to transactions and holdings, respectively. This allows capturing the realities behind PoS networks in a hybrid allocation approach.

---
[2] Transparency note: This article was authored by the founders of CCRI.
[3] Transparency note: This article was co-authored by the founders of CCRI.

The remainder of this white paper is structured as follows: First, we outline guiding principles for designing a framework to account for carbon emissions caused by cryptocurrency and token systems. Second, we define the terminology used in this white paper. Third, we give an overview of different emissions accounting approaches. Fourth, we describe the proposed hybrid allocation approach with its properties and describe how it can be applied in practice to both PoW and PoS. Fifth, we revisit the electricity consumption of Bitcoin and Ethereum (pre- and post-Merge) holdings and transactions by illustratively applying different accounting approaches. Lastly, we conclude and provide avenues for further research.

**Guiding principles for the design of an accounting framework for crypto-related emissions**

We perceive three essential requirements that an accounting framework for crypto-related emissions should satisfy: consistency, continuity, and completeness.

- **Consistency:** We aim for a maximum degree of consistency in the methodology for the underlying systems and their varying parameters. These systems feature different characteristics, e.g., employing different consensus mechanisms (Proof of Work, Proof of Stake, or else), face variable transaction loads, and potentially serve as a base layer for other networks (such as the Lightning network). Furthermore, these features might not be binary (e.g., when do we recognize a transaction volume as significant). We must ensure that the same methodology is applied to similar systems to avoid inconsistent accounting, e.g., by selecting more favorable methodologies for one's own activities.
- **Continuity:** Changing methodologies over time is hard to justify. However, the nature of a system might change in the future. For example, a miner's revenue in Bitcoin primarily stems from the block reward. With future halvings of block rewards, it is assumed that miners receive a larger share of their revenues from transaction fees. If we argue for a methodology change in such cases, it might come as a surprise to entities applying the framework. Also, we would need to define measurable data and thresholds when to switch between methodologies. Potentially, we might have to change the applied methodology frequently for systems with specific characteristics, which is undesirable. Therefore, an accounting framework should aim for lasting application.

---

[4] Transparency note: The industry reports have been commissioned by Avalanche, TRON, Polygon and ConsenSys.

- **Completeness:** We suggest an approach that may cover a wide variety of systems and can account for activity on the application layer and on a second layer. This is specifically important as additional features might get introduced to a system at a later stage (e.g., Cardano recently introduced smart contracts). If applicability to second layer solutions is not ensured, there is a risk of double counting or omitting emissions later.

**Terminology in this paper**

To ensure consistent usage of terms, we briefly define key terminology that is consistently used throughout this paper.

- Layer 1 network: Any network that only relies on its own infrastructure (nodes) for its proper functioning. Example: Bitcoin is a layer 1 network.
- Cryptocurrency: A unit that is the primary means of value or payment within a layer 1 network. Example: Ether is the primary means for paying transaction fees in the layer 1 network Ethereum.
- (Fungible) Token: Tokens are a mean of storing value, rights, or other forms of information within a layer 1 or layer 2 network. They are fungible and can be transferred as desired. Given that they are `within` a layer 1 or layer 2 network and depend on the respective infrastructure, they are entirely dependent on the respective network.
- Non-Fungible Token (NFT): NFTs are tokens that are distinguishable from each other, but otherwise follow the same principles as fungible tokens.
- Token network: Some networks that do not fall strictly under the definition of a layer 1 or layer 2 network, but have a token associated with it, are called token networks. These networks could serve the respective blockchain they are dependent on (given the tokens they leverage). Example: Chainlink provides an oracle service to the blockchain networks they are connected to, while also having a Chainlink token.
- Layer 1/2 application: Any form of smart contract that is running within the respective network. These smart contracts are entirely dependent on the respective network and can manage links to other systems, e.g., an oracle service. Any token or NFT is also a layer 1/2 application. Example: Uniswap provides a decentralized exchange on the Ethereum blockchain.
- Layer 2 network: Any form of network that a) relies partly on its own hardware and infrastructure to keep track of the network state and execute transactions and b) relies partly on a layer 1 network for additional security. Example: Polygon is a layer 2 network, partly relying on the Ethereum network.

**Overview of emissions accounting methods**

While the network-level carbon footprint for the Bitcoin network and increasingly also for other cryptocurrencies has been studied for more than five years, the question of how to allocate carbon emissions to entities with exposure to cryptocurrencies and tokens remains only sparsely covered. Three approaches appear reasonable and feasible:

- **Holding-based:** Distribution of total carbon footprint based on the share of the single holding to the overall market capitalization. For this approach, it is assumed that block validation and mining new coins is the driver for electricity consumption and corresponding emissions. If this assumption is supposed to hold valid, electricity consumption should be (largely) independent of the transactions performed.
- **Transaction-based:** Allocation of total carbon footprint based on the share of transactions. Alternatively, the share of the transaction fee or of transaction size can be utilized. For this approach, it is assumed that transactions are the driver for electricity consumption and corresponding emissions. If this assumption is supposed to hold valid, electricity consumption should fall (close) to zero if no transactions are performed.
- **Hybrid approach:** Combination of the holding-based and the transaction-based method. For PoW networks, based on the split of block reward vs. transaction fees since miners are incentivized by both parts. The share of the total carbon footprint which arises from block rewards is allocated according to the holding-based approach. The share of the carbon footprint incentivized by transactions is allocated accordingly based on rewards due to transaction fees. For PoS networks, the marginal electricity consumption of transactions may serve to identify the share that needs to be allocated to transactions (and holdings, respectively). The shares that are applied to holdings and transactions are network-specific and dynamic over time to account for changes in the underlying drivers.

**Applying the hybrid approach**

Based on the requirements outlined above, a hybrid accounting method appears most promising. It offers a high degree of consistency as no binary decisions (which could be taken tailored in one or the other direction depending on the user's incentive) are required. As a result, cryptocurrencies and tokens with similar incentive structures are treated similarly in the accounting and allocation process. This is also favorable in terms of continuity. Users of the hybrid approach can anticipate how changes in the incentive structure of the protocol will affect the accounting and allocation process since the underlying methodology persists over time. In terms of completeness, the hybrid approach makes it possible to cover a wide range of cryptocurrencies and tokens as it enables to account for the specificities of different systems over an entire spectrum and can also include activity on the application and the second layer.

This section explores key questions to ensure the practical applicability of a hybrid approach that combines holding-based and transaction-based allocation.

### Why should one account for both holdings and transactions?

For both transactions and holdings, claims exist that the respective activity is not contributing to the electricity consumption of the network: With regard to holdings, some claim that the pure act of holding coins and storing this information in memory requires only very low amounts of electricity consumption. Additionally, some coins might have been minted at an early phase of the network in which the mining was comparatively low computationally expensive and thus should not account for the ongoing process of securing the network. With regard to transactions, some claim that the pure act of verifying new transactions is not computationally expensive and even if no transaction occurred the network's electricity consumption might not significantly decrease. To unpack this conflict, it is helpful to understand the drivers of the underlying electricity consumption.

In **Proof of Work** networks, incentivization of the miners to spend money on electricity is a key element. Miners receive a reward for solving a mining puzzle in the form of a block reward and transaction fees from included transactions. For transactions, it is clear that the entity executing the transaction is responsible for the transaction fee (and the thereby resulting incentivization). Consequently, the share of transaction fee in total miner revenue reflects the share in electricity consumption which transactions need to account for. For holdings, it is clear that they massively profit from the ongoing securing of the network as diminished faith in the network would lead to a major price decrease of holdings. Thus, holdings need to account for a certain share of the electricity consumption which should be equally spread across the coins independent of their age to maintain their fungible nature which is an integral part of any currency (coins also equally profit or suffer from price variations). Still, it might remain less obvious why the incentive of the block reward is the responsibility of entities that hold the respective currency. Assuming constant demand for the respective currency, inflation (the creation of new coins) devalues the holdings of holders, implicitly leading to a value transfer of the entity holding cryptocurrency to the miner. Thus, the value of existing holdings and their faith in the network incentivize the miner to participate in the mining process and earn the block reward. In light of the argumentation above, we argue to account for both holdings and transactions, as both contribute to the miner's earnings and, therefore, their incentive; the distribution between block reward and transaction fee is, therefore, a suitable approach.

In **Proof of Stake** networks, the vast majority of electricity consumption is eliminated due to the non-existence of the computationally intensive hash puzzle, but rather only idle as well as the processing of blocks and transactions are sources of electricity consumption. Consequently, emissions from the marginal electricity consumption from transactions should be allocated entirely to entities that execute these transactions, given that they are the reason for the existence of these transactions.[5] The other sources of electricity consumption (i.e., idle of machines and execution of the network client software), we propose to attribute entirely to holdings, as these sources of electricity ensure that the holders of the respective currency can actually spend their coins in exchange for other currencies. If the network stopped running, the respective holdings would be worthless.

**How would the hybrid approach work in practice?**

After identifying which share of the network-level carbon footprint should be allocated to holdings and which share should be allocated to transactions, we need to determine the responsibility in terms of carbon emissions for specific entities. The part of the holding-based approach may be calculated by dividing the holding volume by the overall number of coins of the network at a given time, potentially adjusted for lost coins[6]. Thus, if an entity is holding 1% of all coins of a network, this entity needs to account for 1% of the share that has been identified to be allocated to holdings. For the transaction-based part, we would need to differentiate between PoW and PoS. In PoW, we may aim at the monetary incentives caused by transactions. To do so, we attribute emissions of one transaction to the share of its transaction fees. Thus, if an entity is responsible for 1% of all transaction fees, this entity needs to account for 1% of the share that has been identified to be allocated to transactions. For PoS, we may aim at transaction complexity; that usually stems from a form of gas consumption (e.g., in Ethereum). That seems realistic, as more complex transactions (e.g., executing Smart Contracts) are more computationally expensive than other transactions. If an entity cannot identify its transaction fees or gas consumption, one may also resort to the number of transactions as a second-best proxy. Consequently, if an entity is responsible for 1% of all transactions, this entity needs to account for 1% of the share that has been identified to be allocated to transactions. We suggest not focusing on the monetary value a transaction processes since we are not aware that either in PoW or PoS networks, the value directly influences the incentives for the miner or the computational effort.

**How to include (non-fungible) tokens or layer 1 applications?**

For the question of how to account for (non-fungible) tokens or layer 1 applications, we need to differentiate between PoW and PoS networks. For PoW, a direct holding-based approach for networks such as Ethereum and its tokens appears out of scope, as we do not know the total value of the network. That is because the space moves very fast, the boundaries are unclear, and the value of single objects is hard to measure (e.g., how much are all Cryptopunk NFTs worth?). Thus, we suggest resorting to transactions instead in order to account for these tokens. From here, there are again three ways that could be considered:

- First, applying a transaction-based approach (i.e., only transactions are relevant).
- Second, applying a holding-based approach on a token-related transaction basis. For example, an entity owns 10% of all Uniswap tokens, but does not facilitate any transactions. On a purely transaction-based approach, this entity would not need to account for its holdings; in a token-related transaction approach, the calculation changes: Uniswap is responsible of 50% of the transaction fees on the network; as the entity holds 10% of all Uniswap tokens, it is responsible for 5% of the overall transaction fees (and associated emissions).
- Third, a hybrid approach could be used here by combining the two former approaches weighted by the ratio of block reward and transaction reward of the layer 1 application (e.g., ratio of Ethereum in case of Uniswap).

---

[5] For a detailed description of how the marginal electricity consumption from transactions is derived, please resort to CCRI. (2022a). *Determining the electricity onsumption and carbon footprint of Proof of Stake networks*. https://carbon-ratings.com/dl/whitepaper-pos-methods-2022.

[6] The number of lost coins is estimated to be 20% in Bitcoin. From a technical perspective, lost bitcoins/ether/... can be measured, but no live data is available so far. Data availability would need to be improved over time to include this aspect.

We would also recommend using a hybrid approach, as we would be able to allocate emissions to parties that incentivize other parties to join the respective application. If an overall volume is not available (as the application might not be a token), the purely transaction-based approach may be applied. Combining these methodologies (e.g., hybrid for regular tokens and purely transaction-based allocation for other applications) does not lead to double counting, as the respective base unit for the tokens remains the transaction fee.

For Proof of Stake networks, instead of applying the shares of the block reward and transaction fees, the shares of the base load and the marginal electricity consumption can be leveraged for the hybrid allocation instead.

### How to include layer 2 networks and token networks?

Similar to the approach of allocating carbon emissions to tokens and layer 1 networks, we are also able to allocate emissions to activities on layer 2 networks and token networks. We recommend looking at any activity on the layer 1 network that is because of the layer 2 network and calculate the respective electricity consumption (and carbon footprint, respectively). However, in contrast to a pure layer 1 application, there are additional emissions from the respective infrastructure of the layer 2 network or token network that also need to be accounted for.

If an allocation should take place within a layer 2 network (e.g., what are the emissions of a layer 2 transaction), we suggest to attribute the entire attribution from the layer 1 network to the layer 2 network and apply the hybrid allocation again.

We display a comprehensive overview of the hybrid accounting methodology in Figure 1.

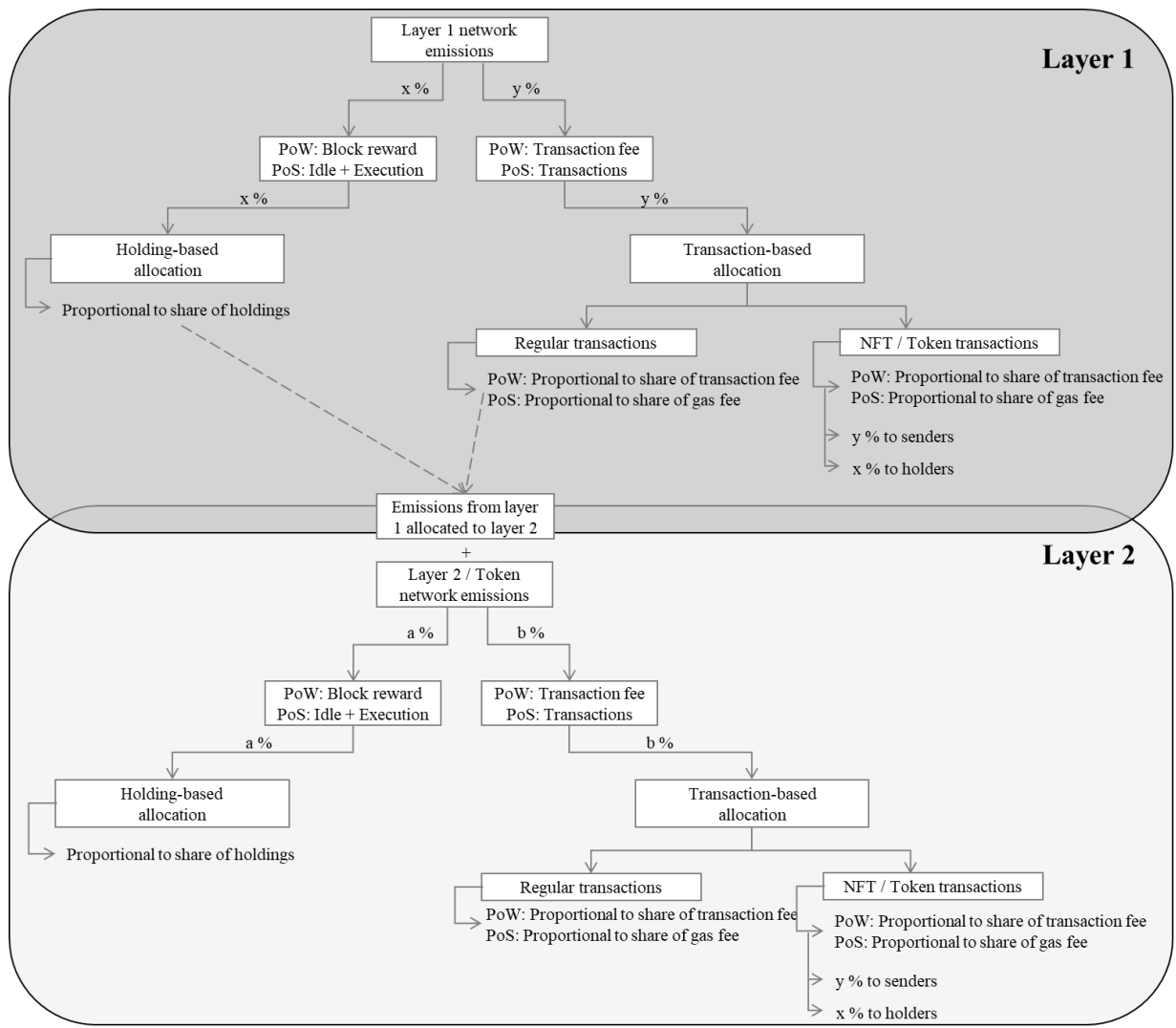

*Figure 1: Hybrid Accounting Methodology for Proof of Work and Proof of Stake networks.*

**The relevance of the hybrid approach in practice: Revisiting the electricity consumption of cryptocurrency transactions**

The relevance of the hybrid approach becomes specifically apparent when looking at calculation examples. There is an ongoing debate about the electricity consumption and carbon footprint of single cryptocurrency transactions, focusing mostly on Bitcoin and Ethereum. For example, Digiconomist[7] claims that one single Bitcoin consumes up to 797.28 kWh of electricity. While the creation, verification and distribution of the transaction in the network hardly utilize any computational power, the number is based on a purely transaction-based approach, dividing all emissions of the network by the amount of transactions in the same time, disregarding the responsibility of cryptocurrency holders and their impact on the carbon footprint of the network.

---

[7] https://digiconomist.net/bitcoin-energy-consumption/, accessed on 13th March 2023

The absence of a framework that accounts both for the responsibility of entities executing transactions and holding cryptocurrency did not allow for a more precise electricity consumption and carbon footprint allocation to transactions. By using the hybrid approach outlined above, we are able to provide a more comprehensive picture of the electricity consumption and carbon footprint of transactions for cryptocurrencies.

### Data sources

We utilize the Cambridge Bitcoin Electricity Consumption Index (CBECI) and the CCRI Crypto Sustainability Indices[8] as a source for the electricity consumption of Bitcoin and Ethereum, respectively. To conduct this analysis, we utilize the CCRI Crypto Sustainability API[9] which allows to also allocate electricity consumption and carbon emission to single holdings and transactions. The Crypto Sustainability API builds on data of the respective blockchains which we obtain by setting up nodes and as well as analyzing block and transaction information from them. For the sake of simplicity, we do not translate the electricity consumption into a carbon footprint for the analysis shown over the remainder of this section.

### Transaction (fee) share as weighting factor

For PoW, the share of transaction fees as a proportion of the overall miner reward determines the distribution of the electricity consumption between holdings and transactions executed. For PoS, we resort to the share of the total network-level power demand that is needed for processing transactions. We select Bitcoin and Ethereum (pre- and post-Merge) as illustrative examples for this analysis, as they show vastly different transaction fee shares as displayed in Figure 2.

---

[8] https://indices.carbon-ratings.com
[9] https://docs.api.carbon-ratings.com

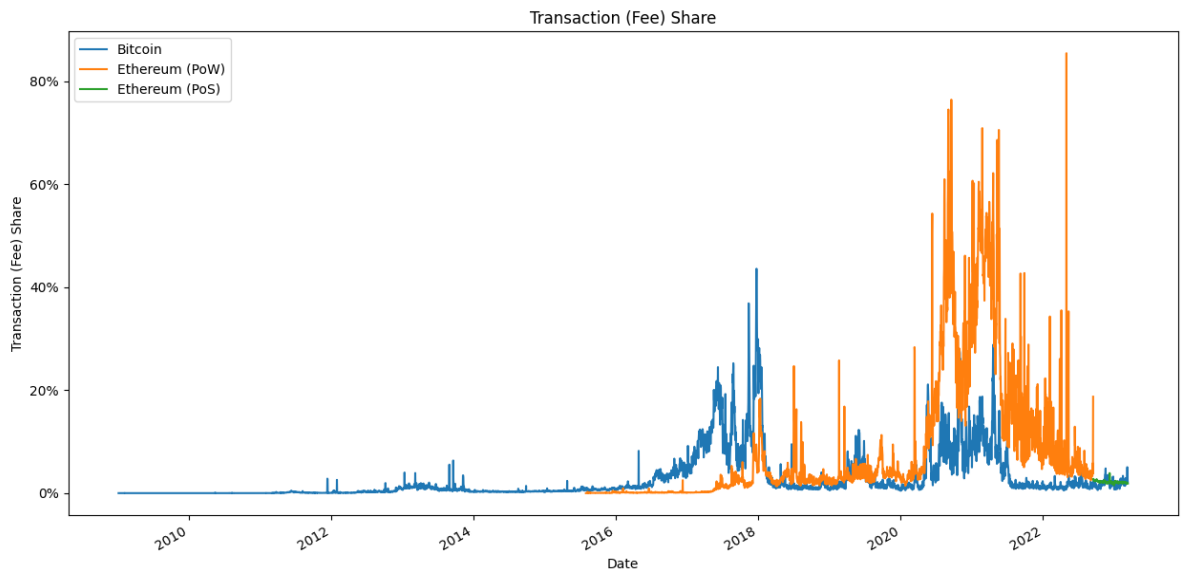

*Figure 2*: *The share of the transaction fee share of the overall miner reward (for Bitcoin and Ethereum PoW) as well as the transaction share of the over all power demand (for Ethereum PoS). Own calculations based on the CCRI Crypto Sustainability API. Ethereum PoW ends and Ethereum PoS starts on 15th September 2022.*

While both networks have seen an increased demand for transactions (and therefore a higher share of the transaction fees), the share has reduced significantly since July 2021. As of March 2023, the transaction fee share for Bitcoin is about 3 %, for Ethereum PoS has about 2 % of transaction share.

These metrics provide clarity on the responsibility of transactions as an incentive for the miner to increase their mining activities: While there have been times in the past in which the transaction fees made up to 70 % of the overall mining reward in Ethereum (summer 2020)[10], these numbers do not warrant to rely on a purely transaction-based metric for the electricity consumption of transactions.

**Electricity allocation**

To derive a more precise estimate for the responsibility of a single transaction within the Bitcoin and Ethereum network, we calculate the electricity consumption for each day in 2021 for executing a bitcoin / ether transaction with a I) holding-based approach, II) transaction-based approach and III) the hybrid approach outlined in this manuscript. For comparison, we also calculate the electricity consumption associated with the holding of one bitcoin / ether for one day with the respective methodologies. Analogously, we also calculate the electricity consumption for each day in Q4 2022 for executing an ether transaction and holding one ether in order to also provide an example for a PoS network. Table 1 displays the values of the respective calculations. The average electricity consumption of the Bitcoin network amounted to 103.57 TWh in 2021, and to 15.59 TWh for Ethereum in 2021. The average transaction fee share for Bitcoin amounts to 6.01 % in 2021, for Ethereum it amounts to 26.6 % in 2021.

---

[10] For Ethereum PoW, there are a few sudden spikes up to 80 % transaction fee share. These are the result of single transactions with very high fees, such as displayed here: https://etherscan.io/txs/label/high-transaction-fee (Accessed 29th January 2023).

For Ethereum PoS, the average transaction share amounts to 2.2 % whereas the average yearly electricity consumption is 2.28 GWh for Q4 2022. Please note that N/A marks that a value is not available, as the respective methodology does not account for the respective activity.

*Table 1: Overview of the daily average electricity consumption allocation in 2021 based on selected methodology to a) holding one coin (bitcoin / ether) for one day and b) performing one transaction. Ethereum PoS data as of 31st December 2022. Own calculations.*

|  | **I) Holding-based** | | **II) Transaction-based** | | **III) Hybrid** | |
| --- | --- | --- | --- | --- | --- | --- |
|  | a) Holding | b) Tx | a) Holding | b) Tx | a) Holding | b) Tx |
| **Bitcoin** | 15.16 kWh | - | - | 1,077.83 kWh | 14.21 kWh | 63.33 kWh |
| **Ethereum (PoW)** | 0.37 kWh | - | - | 33.97 kWh | 0.28 kWh | 7.69 kWh |
| **Ethereum (PoS)** | 0.061 Wh | - | - | 8.41 Wh | 0.060 Wh | 0.149 Wh |

Table 1 highlights the extreme variation in estimates based on the selected allocation method. For Bitcoin, the example is drastic: The electricity consumed by one transaction using a transaction-based allocation is 17 times the number of the electricity consumption using a hybrid approach, even for Ethereum, it is factor 4 in comparison. In contrast, the electricity consumption associated with holding the respective currency for one day using the hybrid approach only drops moderately compared to using a holding-based approach. This is in line with the transaction fee shares shown in Figure 2. Due to the rather low share of transaction fees on average in 2021, a high share of electricity is allocated to holdings in the hybrid approach which consequently leads to a small relative difference between the allocation using the holding-based and the hybrid approach. From these numbers, it is evident that a purely transaction-based approach is unsuitable to properly account for the responsibility of executing a transaction in a blockchain network and therefore should be avoided to be displayed in isolation or to be used for accounting purposes.

**Conclusions**

This white paper presents an approach to account for carbon emissions caused by cryptocurrency and token systems. Here we present an approach that best fits the guiding principles of consistency, continuity, and completeness. Furthermore, the approach appears to be in line with common carbon accounting standards that base the allocation on emission drivers.

Numerous potential allocation approaches have been discussed in the past. Besides the transaction and holding-based approach that form the foundation of the hybrid approach presented here, players in the crypto space have developed further methods. The Green Bitcoin Project, for instance, suggests a model that only accounts for emissions directly linked to the minting of new coins. Others have suggested similar approaches that, for instance, use average electricity consumption to mine new coins since the first coin has been mined (Global Digital Finance, 2021). Such alternative approaches might be beneficial for certain stakeholders, however, appear not compliant with common carbon accounting standards, and might therefore be misleading. Given the outlined data and the high differences in estimates, it becomes of utmost importance to agree on a common accounting methodology.

Irrespective of the chosen approach, data availability remains a key challenge and, therefore, also has implications for approach selection or simplification to ensure applicability. Most research so far has focused on Bitcoin. And even for Bitcoin, results from electricity consumption and carbon emissions chart a wide range of results. However, data availability is improving over time and an increasing number of data points on the electricity consumption of different networks become publicly available. (see for example [CCRI Crypto Sustainability Indices](https://indices.carbon-ratings.com/)[11]).

This white paper only marks the beginning of carbon accounting in the digital asset space. Many questions remain open:

- Derivatives and other financial products might rely on cryptocurrencies and tokens as an underlying asset. How are emissions attributed to these products?
- For many stakeholders along the value chain, emissions from cryptocurrencies and tokens are scope 3 emissions. To which scope 3 emission category should they be allocated (i.e., purchased product and services or investments)? This might also vary across stakeholder groups in the value chain.
- Some institutions, such as stock exchanges, act as pure facilitators for investors and do not have any crypto holdings themselves but contribute to a potential network effect. What is the responsibility of such institutions?

As cryptocurrency and token systems continue their journey into mainstream finance, answers to these questions keep on gaining relevance. Future high-quality research is crucial to avoid stakeholders shaping answers in a way that is beneficial for their business but misleading from a carbon accounting perspective.

---

[11] https://indices.carbon-ratings.com/